# Collaborative Decoding of Interleaved Reed-Solomon Codes using Gaussian Elimination

Hans Kurzweil*, Mathis Seidl† and Johannes B. Huber†
*Department Mathematik, Universität Erlangen-Nürnberg, Erlangen, Germany
†Lehrstuhl für Informationsübertragung, Universität Erlangen-Nürnberg, Erlangen, Germany
Email: kurzweil@mi.uni-erlangen.de, {seidl, huber}@lnt.de

*Abstract*—We propose an alternative method for collaborative decoding of interleaved Reed-Solomon codes. Simulation results for a concatenated coding scheme using polar codes as inner codes are included.

## I. Introduction

Reed-Solomon (RS) codes are used in many applications, often implemented in an interleaved form as outer codes in concatenated code designs. By combining and interleaving several Reed-Solomon codewords, correction of long error bursts affecting only a few symbols of the particular underlying codewords can be achieved.

The standard decoding procedure consists of decoding each of the interleaved codewords separately. In recent years, methods have been investigated which try to decode the columns of the interleaved code no longer independently but in one step, allowing for error correction beyond half the minimum distance.

Our approach is based mainly on the considerations in [1] and [2], involving a significantly smaller linear system of equations when compared to the methods using multi-sequence shift-register synthesis like in [3]. By choosing a parity check matrix of a special form - which we do not have to use explicitly - we obtain an error locator polynomial allowing us to correct errors up to the minimum distance minus 2. The proposed algorithm is of equivalent complexity compared to the multi-sequence approach, but can be easily parallelized in order to increase time performance.

## II. Reed-Solomon codes

A (generalized) Reed-Solomon code of length $n$ and dimension $k$ over a finite field $\mathbb{F}$ with $|\mathbb{F}| = q$ elements can be defined as follows:

**Definition 1** (Reed-Solomon code). *Let $\boldsymbol{v} := (v_1, \ldots, v_n) \in \mathbb{F}^n$ be a vector of $n \leq q$ different elements of $\mathbb{F}$. Let further $\mathcal{P}_k$ be the vector space of polynomials over $\mathbb{F}$ with degree $< k$. Then a Reed-Solomon code $\mathcal{GRS}(q; n, k, \boldsymbol{v})$ is the set of evaluations at $\boldsymbol{v}$*

$$\left\{ (p(v_1), p(v_2), \ldots, p(v_n)) \in \mathbb{F}^n \ : \ p \in \mathcal{P}_k \right\}$$

*of the polynomials from $\mathcal{P}_k$.*

In this work, we will make use of two special types of generalized Reed-Solomon codes:

**Definition 2.** *A Reed-Solomon-Code $\mathcal{GRS}(q; n, k, \boldsymbol{v})$ with length $n = q - 1$ and*

$$\boldsymbol{v} = (\alpha^0, \alpha^1, \ldots, \alpha^{q-2})$$

*with $\alpha$ being a primitive element of $\mathbb{F}$ will be referred to as $\mathcal{RS}(q-1, k)$.*

*The extended code of length $n = q$ obtained by adding the zero element of $\mathbb{F}$ to the vector $\boldsymbol{v}$ of $\mathcal{RS}(q-1, k)$, i.e.*

$$\boldsymbol{v} = (0, \alpha^0, \alpha^1, \ldots, \alpha^{q-2}),$$

*will be called $\mathcal{RS}^*(q, k)$.*

It is known that Reed-Solomon encoding can be accomplished by multiplication with Vandermonde matrices. Therefore, the following matrix is a possible generator matrix of $\mathcal{RS}^*(q, k)$:

$$\boldsymbol{G} = \begin{pmatrix} v_0^0 & v_1^0 & v_2^0 & v_3^0 & \ldots & v_{q-1}^0 \\ v_0^1 & v_1^1 & v_2^1 & v_3^1 & \ldots & v_{q-1}^1 \\ v_0^2 & v_1^2 & v_2^2 & v_3^2 & \ldots & v_{q-1}^2 \\ \vdots & \vdots & \vdots & \vdots & \ddots & \vdots \\ v_0^{k-1} & v_1^{k-1} & v_2^{k-1} & v_3^{k-1} & \ldots & v_{q-1}^{k-1} \end{pmatrix} \quad (1)$$

$$= \begin{pmatrix} 1 & 1 & 1 & 1 & \ldots & 1 \\ 0 & 1 & \alpha & \alpha^2 & \ldots & \alpha^{q-2} \\ 0 & 1 & \alpha^2 & \alpha^4 & \ldots & \alpha^{2(q-2)} \\ \vdots & \vdots & \vdots & \vdots & \ddots & \vdots \\ 0 & 1 & \alpha^{k-1} & \alpha^{2(k-1)} & \ldots & \alpha^{(q-2)(k-1)} \end{pmatrix}. \quad (2)$$

The code $\mathcal{RS}^*(q, k)$ has one interesting property which has been stated in a more general form in [4, p. 304] and will be the foundation of our following considerations:

**Lemma 1.** *The dual code of $\mathcal{RS}^*(q, k)$ is $\mathcal{RS}^*(q, q-k)$, i.e. the Vandermonde matrix*

$$\boldsymbol{H} = \begin{pmatrix} 1 & 1 & 1 & 1 & \ldots & 1 \\ 0 & 1 & \alpha & \alpha^2 & \ldots & \alpha^{q-2} \\ 0 & 1 & \alpha^2 & \alpha^4 & \ldots & \alpha^{2(q-2)} \\ \vdots & \vdots & \vdots & \vdots & \ddots & \vdots \\ 0 & 1 & \alpha^{m-1} & \alpha^{2(m-1)} & \ldots & \alpha^{(q-2)(m-1)} \end{pmatrix}$$

*with $m := n-k$ is a possible parity check matrix of $\mathcal{RS}^*(q, k)$.*

*Proof:* As can easily be observed, each element of the matrix product $\boldsymbol{G}\boldsymbol{H}^\top$ except for the upper left element $(\boldsymbol{G}\boldsymbol{H}^\top)_{11}$

takes the form

$$a_{mn} := (\boldsymbol{GH}^\top)_{mn} = \sum_{i=0}^{q-2} x^i$$

for some $x \in \mathbb{F} \setminus \{0, 1\}$. Clearly,

$$x^{q-1} = x^0 = 1$$

holds for all $x \in \mathbb{F}$, $x \neq 0$. Consequently, we obtain

$$x \cdot a_{mn} = \sum_{i=0}^{q-2} x^{i+1} = \sum_{i=1}^{q-1} x^i = a_{mn}$$

and because of $x \neq 1$,

$$a_{mn} = 0$$

follows. (we remark that the discrete Fourier transformation is based on the very same argument). The residual matrix element $a_{11}$ simply consists of a sum of ones which adds up to zero:

$$(\boldsymbol{GH}^\top)_{11} = \sum_{i=1}^{q} 1 = 0,$$

as the order of the additive group of $\mathbb{F}$ is $q$.
Therefore, $\boldsymbol{GH}^\top = \boldsymbol{0}$. ∎

## III. INTERLEAVED REED-SOLOMON CODES AND COLLABORATIVE DECODING

By grouping $l \in \mathbb{N}$ codewords of $\mathcal{GRS}(q; n, k, \boldsymbol{v})$ to a matrix, we obtain a linear code of length $(l \cdot n)$, dimension $(l \cdot k)$ and minimum distance like the individual columns $(n-k+1)$.

**Definition 3** (**Interleaved Reed-Solomon (IRS) code**). *Given a certain Reed-Solomon code $\mathcal{C} := \mathcal{GRS}(q; n, k, \boldsymbol{v})$, we define an Interleaved Reed-Solomon code $\mathcal{IRS}(q; l, n, k, \boldsymbol{v})$ of interleaving depth $l$ as the set of $(n, l)$-matrices*

$$\left\{ \boldsymbol{A} = \left(\boldsymbol{a}^{(1)}, \boldsymbol{a}^{(2)}, \ldots, \boldsymbol{a}^{(l)}\right) \; : \; (\boldsymbol{a}^{(i)})^\top \in \mathcal{C}, \; i = 1, \ldots, l \right\},$$

*each consisting of $l$ column-wise arranged codewords from $\mathcal{C}$. In case of $\mathcal{C} = \mathcal{RS}^*(q, k)$, the resulting IRS code will be referred to as $\mathcal{IRS}^*(q, l, k)$.*

Assume that a codeword $\boldsymbol{A} \in \mathcal{IRS}^*(q, l, k)$ is transmitted over a noisy channel, so that

$$\boldsymbol{Y} = \boldsymbol{A} + \boldsymbol{E} \; \in \mathbb{F}^{n \times l}$$

with some error matrix $\boldsymbol{E} \in \mathbb{F}^{n \times l}$ is received at the channel output. Let $\boldsymbol{E}$ be a matrix with exactly $f \in \mathbb{N}$ non-zero rows. We denote $\mathcal{F}$ the set of indices of these erroneous rows.
As each column of $\mathcal{IRS}^*(q, l, k)$ is a codeword of $\mathcal{RS}^*(q, k)$, we could try to decode each of the $l$ columns separately by computing a syndrome sequence $\boldsymbol{s}^{(i)} := (s_1^{(i)}, \ldots, s_{n-k}^{(i)})$ of length $(n-k)$ and solving a system of linear equations for each $i \in \{1, \ldots, l\}$ (which can be done efficiently by algorithms like the well-known Berlekamp-Massey-Algorithm in $\mathcal{O}(f^2)$). By this means we would be able to correct up to $t := \lfloor \frac{n-k}{2} \rfloor$ errors per column which corresponds to half of the minimum distance of $\mathcal{RS}^*(q, k)$.

For collaborative decoding, we arrange the $l$ syndrome sequences of an IRS codeword as columns of a so-called syndrome matrix $\boldsymbol{S}$. The computation can be written formally as a matrix multiplication of $\boldsymbol{Y}$ with the parity check matrix $\boldsymbol{H}$ of the underlying $\mathcal{RS}^*$-Code:

$$\boldsymbol{S} = \boldsymbol{H} \cdot \boldsymbol{Y} = \boldsymbol{H} \cdot (\boldsymbol{A} + \boldsymbol{E}) = \boldsymbol{H} \cdot \boldsymbol{E} = \boldsymbol{H}^\mathcal{F} \cdot \boldsymbol{E}_\mathcal{F} \quad (3)$$

with $\boldsymbol{H}^\mathcal{F}$ and $\boldsymbol{E}_\mathcal{F}$ denoting the submatrices of $\boldsymbol{H}$ and $\boldsymbol{E}$ consisting only of those columns and lines, respectively, whose indices lie in $\mathcal{F}$. The last equivalence holds because all other rows of $\boldsymbol{E}$ are zero. Note that according to Lemma 1 the computation of $\boldsymbol{S}$ consists plainly of polynomial evaluations. The syndrome matrix takes the form

$$\boldsymbol{S} = \begin{pmatrix} s_1^{(1)} & s_1^{(2)} & \cdots & s_1^{(l)} \\ s_2^{(1)} & s_2^{(2)} & \cdots & s_2^{(l)} \\ \vdots & \vdots & \ddots & \vdots \\ s_{n-k}^{(1)} & s_{n-k}^{(2)} & \cdots & s_{n-k}^{(l)} \end{pmatrix} \in \mathbb{F}^{(n-k) \times l}. \quad (4)$$

Our decoding algorithm succeeds whenever the following two conditions are fulfilled:

$\mathcal{H}_1$ The non-zero rows $\boldsymbol{E}_i$ ($i \in \mathcal{F}$) of the received word are linear independent.
$\mathcal{H}_2$ $f \leq f_{\max} := \min\{l, n-k-1\}$ holds.

Actually, $\mathcal{H}_1$ and $\mathcal{H}_2$ are both necessary and sufficient for successful decoding. We will make some remarks on the linear independence condition $\mathcal{H}_1$ later in section VI.
The square submatrix of $\boldsymbol{H}$

$$\boldsymbol{K} := \left(\boldsymbol{H}^\mathcal{F}\right)_{[f]} \in \mathbb{F}^{f \times f} \quad (5)$$

consisting of the first $f$ rows of $\boldsymbol{H}^\mathcal{F}$ is a Vandermonde and thus non-singular matrix. Therefore, the following line of $\boldsymbol{K}$

$$\boldsymbol{\mu} := \left(\boldsymbol{H}^\mathcal{F}\right)_{f+1} \in \mathbb{F}^f \quad (6)$$

(note $f < n-k$) is an uniquely determined linear combination of the first $f$ rows $\boldsymbol{K_j}$ of $\boldsymbol{K}$:

$$\boldsymbol{\mu} = \sum_{j=1}^{f} \lambda_j \boldsymbol{K}_j \quad (7)$$

for some $\lambda_j \in \mathbb{F}$, $j = 1, \ldots, f$. Given these coefficients, we define a polynomial

$$\Lambda(x) := x^f - \sum_{j=1}^{f} \lambda_j x^{j-1}. \quad (8)$$

Due to the special form of $\boldsymbol{H}$ (cf. Lemma 1), the $i$th column $\boldsymbol{H}^i$ consists of the consecutive powers of $v_i \in \mathbb{F}$. Consequently,

$$0 = \Lambda(x_i) = x_i^f - \sum_{j=1}^{f} \lambda_j x_i^{j-1} \quad (9)$$

holds for all $i \in \mathcal{F}$. Since $\Lambda$ is a polynomial of degree $f = |\mathcal{F}|$, these are obviously the only roots. Therefore, $\Lambda$ is the error locator polynomial. In the following, we will determine the coefficients $\lambda_j$ from eq. (3):
Let $\varphi$ be the linear mapping defined by $\boldsymbol{E}_\mathcal{F} \in \mathbb{F}^{f \times l}$:

$$\varphi \; : \; \mathbb{F}^f \mapsto \mathbb{F}^l \quad , \quad x \mapsto x \cdot \boldsymbol{E}_\mathcal{F} \quad (10)$$

By definition,

$$\varphi(\boldsymbol{K}_i) = \boldsymbol{S}_i \quad , \quad i = 1, \ldots, f \qquad (11)$$

By our assumption $\mathcal{H}_1$, $\boldsymbol{E}_{\mathcal{F}}$ is a matrix of maximum rank $f$. Thus, $\varphi$ is an injective mapping, and the vectors $\boldsymbol{S}_i$ ($i = 1, \ldots, f$) form a basis of the image of $\varphi$. Consequently, the $(f+1)$th row of $\boldsymbol{S}$ is a linear combination of the former, uniquely determined by the coefficients of $\Lambda$:

$$\boldsymbol{S}_{f+1} = \varphi(\boldsymbol{\mu}) = \sum_{j=1}^{f} \lambda_j \boldsymbol{S}_j. \qquad (12)$$

These coefficients can now be calculated by performing elementary column operations and determining a minimum index $\hat{f}$ such that the $(\hat{f}+1)$-th row is a linear combination of the preceding rows $\boldsymbol{S}_1, \ldots, \boldsymbol{S}_{\hat{f}}$ of the syndrome matrix $\boldsymbol{S}$. By applying the Gauss-Jordan algorithm to $\boldsymbol{S}$, we obtain a matrix of the form

$$\tilde{\boldsymbol{S}} := \begin{pmatrix} 1 & 0 & \ldots & 0 & 0 & \ldots & 0 \\ 0 & 1 & \ldots & 0 & 0 & \ldots & 0 \\ \vdots & \vdots & \ddots & \vdots & \vdots & \ddots & 0 \\ 0 & 0 & \ldots & 1 & 0 & \ldots & 0 \\ \lambda_1 & \lambda_2 & \ldots & \lambda_{\hat{f}} & 0 & \ldots & 0 \\ \vdots & \vdots & \vdots & \vdots & \vdots & \vdots & \vdots \end{pmatrix}. \qquad (13)$$

Under the assumption $\mathcal{H}_1$ such an index $\hat{f}$ surely exists as the rank of $\boldsymbol{S}$ is smaller than the number of rows $(n-k)$.

As a consequence, we obtain

**Theorem 1.** *If $\mathcal{H}_1$ and $\mathcal{H}_2$ are fulfilled, then $\hat{f} = f$ and the polynomial*

$$\Lambda(x) = x^{\hat{f}} - \sum_{j=1}^{\hat{f}} \lambda_j x^{j-1}$$

*built from the elements $\lambda_j$ from (13) is the error locator polynomial, i.e.*

$$\Lambda(v_i) = 0 \quad \Leftrightarrow \quad i \in \mathcal{F}$$

*holds.*

The set of error locations $\mathcal{F}$ can now be determined by finding the roots of $\Lambda(x)$ using standard methods like the Chien search.

Obviously, in principle only $(f+1)$ columns of $\boldsymbol{S}$ are necessary in order to determine $\Lambda$, allowing for significant reductions in computational complexity. We will dwell on this possibility in section VII.

## IV. Codeword reconstruction

Given the (correctly computed) set $\mathcal{F}$ of erroneous columns of $\boldsymbol{Y}$, we are now able to reconstruct $\boldsymbol{E}_{\mathcal{F}}$ and therefore $\boldsymbol{A} = \boldsymbol{Y} - \boldsymbol{E}$. The matrix equation

$$\boldsymbol{S} = \boldsymbol{H}^{\mathcal{F}} \cdot \boldsymbol{E}_{\mathcal{F}} \qquad (14)$$

defines an over-determined system of linear equations consisting of $l \cdot (n-k)$ equations and $l \cdot f$ unknowns. Since we know that (14) must have an unique solution and since the rows of the (Vandermonde!) matrix $\boldsymbol{H}^{\mathcal{F}}$ are linear independent, we can restrict to the smaller system

$$\boldsymbol{S}_{[f]} = \boldsymbol{H}^{\mathcal{F}}_{[f]} \cdot \boldsymbol{E}_{\mathcal{F}} = \boldsymbol{K} \cdot \boldsymbol{E}_{\mathcal{F}} \qquad (15)$$

with $\boldsymbol{S}_{[f]}$ and $\boldsymbol{H}^{\mathcal{F}}_{[f]}$ denoting the matrices consisting of the first $f$ rows of $\boldsymbol{S}$ and $\boldsymbol{H}^{\mathcal{F}}$, respectively.

As mentioned before, $\boldsymbol{K}$ is a quadratic - and thus invertible - Vandermonde matrix and (15) actually an interpolation problem.

## V. Cyclic and shortened codes

It is well known that using generalized, non-cyclic Reed-Solomon codes leads to an increased computational complexity of the encoding procedure as well as of the syndrome calculations. In our case, by skipping the first element $v_0 = 0$ of $\boldsymbol{v}$ we obtain a cyclic code of length $n = q-1$ which can be efficiently treated by use of the discrete Fourier transformation (DFT) or a generator polynomial. To preserve the duality property of Lemma 1 (which is crucial for our algorithm to work), we treat the evaluations at the skipped element - which we do not know - as a "dummy" row of the IRS codeword. Consequently, we only transmit the rows with indices 2 to $q$.

On the receiver side we assume an all-zero first row. Therefore, we can again use DFT for the computation of the syndromes. Certainly, the correct first row (obtained by evaluation of the information polynomials at $v_1 = 0$) will most probably not be all-zero. In this case, the error locator polynomial $\Lambda$ will simply have an additional root at $v_1 = 0$, causing a shift of the coefficients by one. Otherwise, the constant coefficients of the information polynomials are in fact equal to zero, which has no effect on $\Lambda$.

Both the maximum number of errors we are able to correct and the (actual used) code length decrease by one; the performance remains unchanged.

Further shortening may be applied by the usual methods in order to obtain greater flexibility for code design.

## VI. Failure and error probability

As shown before, in case of $f \leq f_{\max}$ the success of the decoding procedure solely depends on the linear independence of the error vectors. If we assume that the $\boldsymbol{E}_i$ are random vectors uniformly distributed over $\mathbb{F}^l \setminus \{\boldsymbol{0}\}$, the probability for the $\boldsymbol{E}_i$ being dependent, i.e. the probability that $\mathcal{H}_1$ is not fulfilled, can be overbounded for $f \geq 2$ by

$$q^{-(l+1-f)} \cdot \frac{1 - q^{-f}}{1 - q^{-1}} \approx q^{-(l+1-f)}, \qquad (16)$$

as shown in [1]. Clearly, the decoder certainly fails if the number of erroneous columns exceeds $f_{\max}$. Thus,

$$P_f(f, l) \leq \begin{cases} 0 & f < 2 \\ q^{-(l+1-f)} & 2 \leq f \leq f_{\max} \\ 1 & \text{else} \end{cases} \qquad (17)$$

holds as an upper bound for the failure probability under assumption of uniformly distributed error vectors.

If the rows of $\boldsymbol{E}_{\mathcal{F}}$ are linear dependent, the decoder either is not able to obtain any result or decides for a wrong codeword. In the following, we will derive an upper bound for the second case:

Looking again at the linear mapping $\varphi$ from the considerations preceding Theorem 1, there are now some non-zero vectors $\boldsymbol{u} \in \mathbb{F}^f$ - which represent linear combinations of the rows of $\boldsymbol{E}_\mathcal{F}$ - with $\varphi(\boldsymbol{u}) = \boldsymbol{0} \in \mathbb{F}^l$. As the (now linear dependent) error vectors $\boldsymbol{E}_i$ are random vectors, each vector from $\mathbb{F}^f \setminus \{\boldsymbol{0}\}$ can be assumed to have equal probability to lie in the kernel of $\varphi$.

Because of the linear dependence together with $\mathcal{H}_2$, the rank $t$ of $\boldsymbol{S}$ is certainly smaller than $(n-k)$, and therefore the algorithm in any case determines a polynomial $\Lambda$ of degree $t$. If this polynomial has exactly $t$ different roots, we will call it a $t$-valid polynomial, in accordance to [3]. If $\Lambda$ is not $t$-valid, it is obvious that it cannot be the wanted error locator polynomial. In this case the algorithm reports a decoding failure without taking any decision. The fraction of $t$-valid polynomials with leading coefficient 1 and degree $t$ is given by

$$P_v(t) := \binom{q}{t} \cdot q^{-t} \leq \frac{1}{t!} \ . \quad (18)$$

If the algorithm determines a minimal zero combination of the first rows of $\boldsymbol{S}$ which define a linear subspace of smaller dimension than the rank of $\boldsymbol{S}$, the failure is also detected. Therefore, for an erroneous result the first $t < f$ vectors can be assumed to be linear independent while any other error vector has to be a linear combination of the preceding.

Consequently, the probability for obtaining a wrong result can be overbounded by

$$P_e(f,l) \leq \sum_{t=2}^{f-1} \left( P_v(t) \cdot \left(1 - P_f(t,l)\right) \cdot \prod_{j=t}^{f-1} q^{t-l} \right) \quad (19)$$

$$\leq \sum_{t=2}^{f-1} P_v(t) \cdot q^{-(l-t)(f-t)} \quad (20)$$

$$\leq \sum_{t=2}^{f-1} \frac{1}{t!} \cdot q^{-(l-t)(f-t)}. \quad (21)$$

Since the sum is clearly dominated by its last summand,

$$P_e(f,l) \approx \frac{1}{(f-1)!} \cdot q^{-(l+1-f)} \quad (22)$$

holds approximately. Note that this probability is already included in $P_f(f,l)$.

Compared with [3], the probabilities for the decoding to fail are equivalent in case of the maximum number of correctable errors $f = f_{\max} = (n-k-1)$, but declines slower with decreasing $f$. Still, both $P_f$ and $P_e$ fall exponentially with increasing interleaving depth $l$ and thus can be designed arbitrary small by moderate expense.

In case of concatenated code designs where the columns of an outer IRS code are encoded by an inner block code, the probability for a column of an IRS codeword to become corrupted is given by the frame error rate $p_i$ of the inner code. The overall frame error rate (FER) with respect to $p_i$ can then be analytically determined by

$$\text{FER} = \sum_{t=2}^{N} \binom{N}{t} \cdot P_f(t,l) \cdot p_i^t \cdot (1-p_i)^{N-t}. \quad (23)$$

The probability that the decoder decides for a wrong codeword is derived in an analogous manner:

$$\text{FER}_\text{e} = \sum_{t=2}^{N} \binom{N}{t} \cdot P_e(t,l) \cdot p_i^t \cdot (1-p_i)^{N-t}. \quad (24)$$

Fig. 1 depicts the bounds on the FER as well as on the error probability $\text{FER}_\text{e}$ as a function of the inner code error rate $p_i$. Here, a $(204, 188)$ shortened RS code like in the DVB standard [7] is used. As the corresponding IRS code is able to correct up to 15 erroneous columns, we chose an interleaving depth of $l = 16$ rather than $l = 12$ in the DVB standard. Fig. 1 demonstrates that the probability for taking a wrong decision should be usually negligible compared to the failure probability. The bound considered in the multi-sequence approach (cf. [3]) coincides with the FER bound derived here.

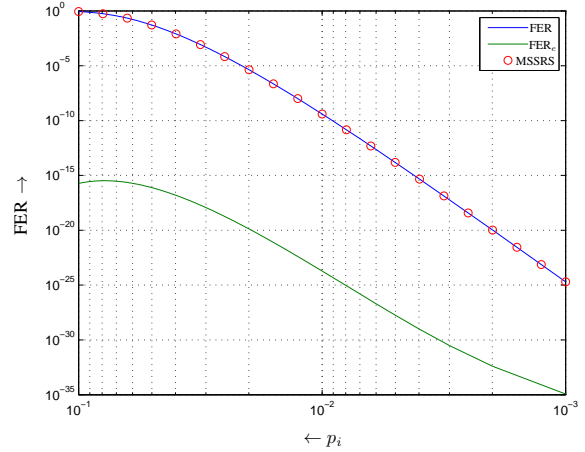

Fig. 1. Bounds FER and $\text{FER}_\text{e}$ for a $(204, 188)$ IRS code with $l = 16$ for collaborative decoding using Gaussian elimination and failure bound for collaborative decoding using multi-sequence shift-register synthesis (MSSRS)

We remark that further increasing of the interleaving depth $l$ has practically no effect on the failure performance which is dominated by the probability that the number of erroneous rows exceeds $f_{\max}$.

## VII. COMPLEXITY

As the system of linear equations considered in this paper has no Toeplitz structure in contrast to the single decoding case, we have to use Gaussian elimination, an algorithm known to have cubic time complexity. In our case of rectangular matrices the complexity is of order $\mathcal{O}(l \cdot f^2)$ (with $l$ and $f$ as defined before).

However, only a single system has to be solved whereas there are $l$ systems in the independent decoding case as well as in [3]. Moreover, in each step of the Gaussian elimination, the $l$ columns of the syndrome matrix $\boldsymbol{S}$ may be transformed at the same time by a parallel implementation of the decoder. Such parallelized algorithms for Gaussian elimination are able to reduce the time complexity up to $\mathcal{O}(l)$.

As mentioned before, rather than the whole matrix $\boldsymbol{S}$ only the first $(f+1)$ columns are actually necessary for determination of the error locator polynomial. In the case of $f \ll l$ it is therefore possible to reduce computational complexity, applying a decoding algorithm which successively performs Gaussian elimination on quadratic submatrices of $\boldsymbol{S}$ of increasing size whereas in the $i$th step the $(i \times i)$-submatrix consisting of the first $i$ rows and columns of $\boldsymbol{S}$ is considered. Decoding stops when a submatrix with linear dependent rows is found. The linear combination obtained in this way may now be "tested" on a certain number of additional columns in order to ensure the desired level of reliability.

## VIII. SIMULATION RESULTS

Finally, we present some examplary simulations to demonstrate the tightness of the derived bounds and to show that the randomness of the error vectors can be realized in practice. For simulation, a (256,128) polar code as inner code is used

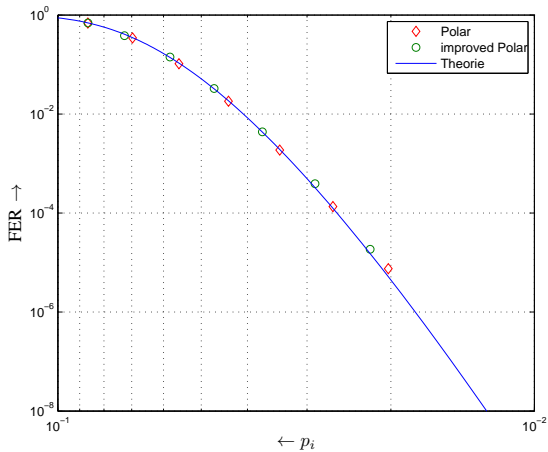

Fig. 2. Simulated and analytical FER of a concatenated code composed of an inner $(256, 128)$ polar code and an outer RS(204,188) code with $l = 16$ as a function of inner code FER

together with the (204,188) IRS code from section VI. Polar codes [5], first introduced by E. Arıkan, are decoded by a low-complexity successive decoder which generates estimations on the source bits one after another, each depending on the decisions made before. In case of a wrong decision, from that bit on unpredictable, long error sequences are produced until the end of the codeword. This fact (which could usually be seen as a drawback) makes polar codes well suited as inner codes in our case. However, the polar successive decoder happens to fail at certain bits significantly more likely than at other ones which prevents the error vectors from being uniformly distributed. It turned out be necessary to apply different bit permutations for each column of the IRS codeword before encoding with the inner polar code in order to randomize the error vectors at the RS decoder input.

Fig. 2 depicts the FER of the concatenated block as a function of the word error rate of the inner polar code, transmitted over the binary-input AWGN channel. The diamond

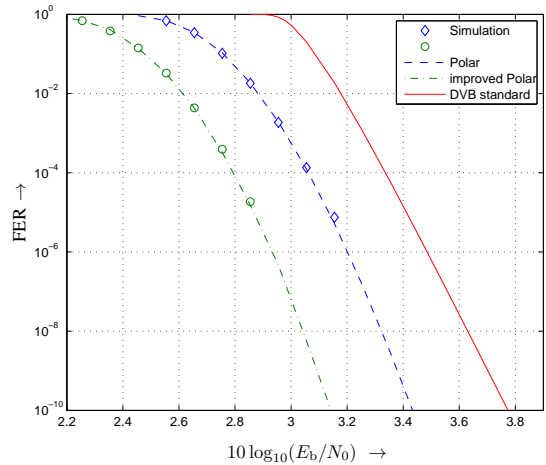

Fig. 3. Simulated and analytical FER performance of a $l = 16$ concatenated code with an outer RS(204,188) code and (a) an inner (256,128) polar code with standard decoding, (b) an inner (256,128) polar code with improved decoding, (c) an inner rate 1/2 convolutional code with constraint length $K = 7$

markers correspond to simulation points where $E_b/N_0 = 2.85\,\mathrm{dB}\ldots3.15\,\mathrm{dB}$ with a step size of $0.1\,\mathrm{dB}$ while the continuous line demarks the analytically derived bound from section VI. By using an improved polar decoding scheme as considered in [6], the performance can be further enhanced by about $0.3\,\mathrm{dB}$. Simulation results of those improved polar codes are represented by circular markers. Fig. 2 shows that the simulation points of both decoding schemes match the theoretical failure bound.

In Fig. 3 the FER performance of the two polar coding schemes are compared to a concatenation of an inner rate $1/2$ convolutional code (constraint length $K = 7$) with an independently decoded outer $(204, 188)$ RS code as used in the DVB standard, but for $l = 16$ instead of $l = 12$.

Both IRS-polar concatenation schemes clearly outperform the DVB code in terms of frame error rates as well as of computational complexity.